\UseRawInputEncoding
\documentclass[aip,reprint,floatfix,superscriptaddress]{revtex4-2}

\usepackage{amsmath}
\usepackage{amssymb}
\usepackage{graphicx}
\usepackage{dcolumn}
\usepackage{natbib}
\usepackage{xr}
\usepackage{cleveref}
\usepackage{bm}
\usepackage{xcolor}
\usepackage{moreverb} 

\usepackage{verbatim}

\providecommand{\keywords}[1]
{
  \small	
  \textbf{\textit{Keywords---}} #1
}

\begin{document}

\newcommand{\BU}{Department of Mechanical Engineering, Division of Materials Science and Engineering, and the Photonics Center, Boston University, Boston, Massachusetts 02215, USA}

\title{Dynamics of  NEMS Resonators across Dissipation Limits}

\author{C. Ti}
\affiliation{\BU}

\author{J. G. McDaniel}
\affiliation{\BU}

\author{A. Liem}
\affiliation{\BU}

\author{H. Gress}
\affiliation{\BU}

\author{{M. Ma}}
\affiliation{\BU}

\author{S. Kyoung}
\affiliation{\BU}

\author{O. Svitelskiy}
\affiliation{Department of Physics, Gordon College, Wenham, Massachusetts 01984, USA}

\author{C. Yanik}
\affiliation{SUNUM, Nanotechnology Research and Application Center, Sabanci University, Istanbul, 34956, Turkey}

\author{I. I. Kaya}
\affiliation{SUNUM, Nanotechnology Research and Application Center, Sabanci University, Istanbul, 34956, Turkey}
\affiliation{Faculty of Engineering and Natural Sciences, Sabanci University, Istanbul, 34956, Turkey}

\author{M. S. Hanay}
\affiliation{Department of Mechanical Engineering, Bilkent University, Ankara, 06800, Turkey}
\affiliation{UNAM -- Institute of Materials Science and Nanotechnology, Bilkent University, Ankara, 06800, Turkey}

\author{M. Gonz\'alez}
\affiliation{Aramco Americas, Aramco Research Center--Houston, Houston, Texas 77084, USA}

\author{K. L. Ekinci}
\email[Electronic mail: ]{ekinci@bu.edu}
\affiliation{\BU}


\date{\today}


\begin{abstract}
The oscillatory dynamics of nanoelectromechanical systems (NEMS) is at the heart of many emerging applications in nanotechnology. For  common   NEMS, such as  beams and strings,  the oscillatory  dynamics is formulated using a dissipationless wave equation  derived from elasticity. Under a  harmonic ansatz, the wave equation gives an undamped free vibration equation; solving this equation with the proper boundary conditions provides the undamped eigenfunctions with the familiar standing wave patterns. Any harmonically driven solution is expressible in terms of these undamped eigenfunctions. Here, we  show that this formalism becomes  inconvenient as  dissipation increases.  To this end, we experimentally map out the position- and frequency-dependent oscillatory motion  of a  NEMS string resonator  driven linearly by a non-symmetric force at one end at different dissipation limits.  At low dissipation (high $Q$ factor), we observe sharp resonances  with standing wave patterns  that closely match the eigenfunctions of an undamped string. With  a slight increase in dissipation, the standing wave patterns become lost and  waves begin to propagate along the nanostructure.  At large dissipation (low $Q$ factor), these propagating waves  become strongly attenuated and display little, if any, resemblance to the undamped string eigenfunctions.  A more efficient and intuitive description of the oscillatory dynamics of a NEMS resonator can be obtained by superposition of waves propagating  along the nanostructure.
\end{abstract}

\maketitle


Nanoelectromechanical systems (NEMS) have enabled a number of nanotechnologies for monitoring the environment \cite{bargatin2012large}, storing and processing information \cite{fan2016integrated, rabl2010quantum}, and applying controllable forces to physical \cite{sohn2018controlling} and biological nanosystems \cite{marshall2003direct,raman2011mapping}. NEMS-based detection of individual atoms and molecules \cite{jensen2008atomic, hanay2012single}, single charge quanta \cite{cleland1998nanometre, lahaye2009nanomechanical}, and vibrations of single microorganisms \cite{gil2020optomechanical,malvar2016mass} has established the potential of NEMS sensors. NEMS are also at the forefront of fundamental physical science, opening up studies in quantum mechanics \cite{o2010quantum}, optomechanics \cite{anetsberger2009near}, Brownian motion \cite{ari2020nanomechanical,maillet2017nonlinear}, fluid mechanics \cite{kara2017generalized,fong2019phonon,liem2021acoustic,liem2021nanoflows}, and nanoelectronics \cite{ti2021frequency, low2012electron, khivrich2019nanomechanical,sapmaz2006tunneling}.


\begin{figure*}[ht!]
    \includegraphics[width=6.75in]{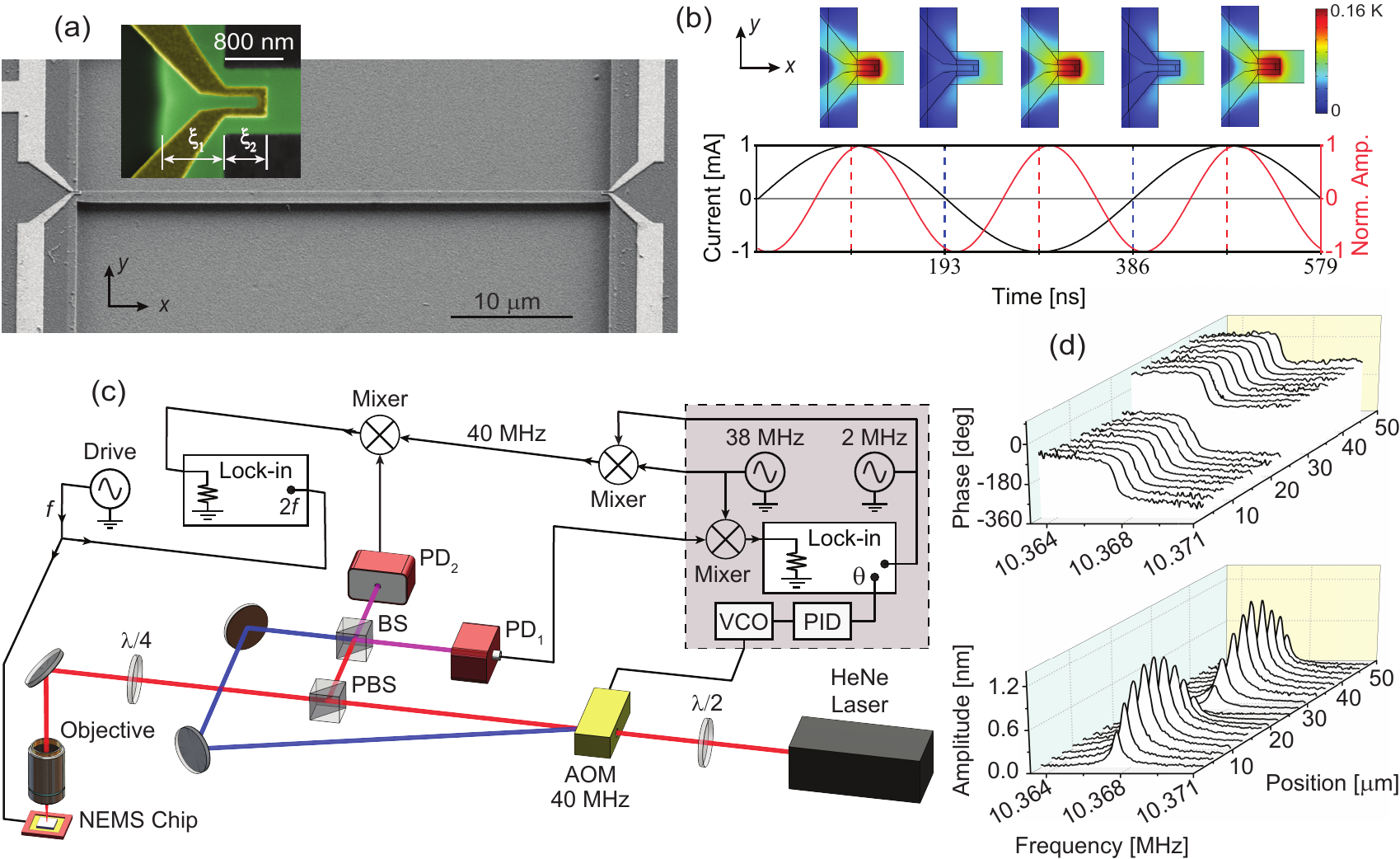}
    \caption{(a) SEM image of a tension-dominated silicon nitride doubly-clamped beam with linear dimensions (along $x$, $y$, $z$) of $l\times b \times h \approx 50~ {\rm \mu m} \times 900~{\rm nm} \times 100~{\rm nm}$. The two identical u-shaped gold  thin-film nanoresistors of thickness of 135 nm and  width of 120 nm on the  anchors of the  NEMS act as    electrothermal actuators. The dimensions along $x$ are $\xi_1 \approx 800$ nm and  $\xi_2 \approx 600$ nm. (b) Numerical simulations of  electrothermal actuation near (but \textit{not exactly} at) the fundamental resonance frequency in vacuum.   Sinusoidal current input to the nanoresistor at $f = 2.5895$ MHz (black curve) results in  nanomechanical  oscillations of the beam  at $f = 5.179$ MHz. The response of the beam at its center is shown. The small phase between the current and displacement is  due to  the thermal inertia and the mechanical response of the resonator.  The  color maps  are the temperature profiles  of the region in inset of (a) at five instants.  Typical  powers dissipated on the nanoresistor are 1 $\mu$W, 50 $\mu$W and 100 $\mu$W in vacuum, air, and water, respectively. (c) Heterodyne optical interferometer. AOM: acousto-optic modulator; $\lambda/2$: half wave plate; $\lambda/4$: quarter wave plate; PBS: polarizing beam splitter; BS: beam splitter; PD: photodetector; PID: proportional-integral-derivative controller; VCO: voltage controlled oscillator. The signal on $\rm PD_1$ is used for feedback; $\rm PD_2$ is connected to a lock-in amplifier via a mixer for driven measurements.  The lock-in amplifier is used in $2f$ mode. (d) Phase (top) and amplitude for a beam  as  functions of frequency and position at its first harmonic resonance in vacuum.}
    \label{Figure1}
\end{figure*}

In a typical  implementation \cite{kouh2017nanomechanical}, one actuates linear  oscillations of the NEMS resonator using a force transducer and looks for  changes in the phase, frequency, or dissipation due to interactions. For proper operation, the user must know  how exactly the nanomechanical structure  is moving under the actuation forces.  The oscillatory NEMS dynamics is typically determined using a \textit{dissipationless} wave equation, e.g., the beam equation or the string equation, derived from elasticity. After  the harmonic ansatz, one obtains the \textit{undamped} free vibration equation and solves it subject to  boundary conditions \cite{cleland2002noise, cleland2013foundations}. This approach provides the \textit{undamped} eigenfunctions that correspond to standing wave patterns on the structure. These well-known patterns emerge from the interference of undamped waves reflecting back and forth from the boundaries of the structure.  The undamped eigenfunctions form a complete set, and the driven harmonic motion of NEMS, even in the presence of dissipation, can be expressed as an expansion in terms of these eigenfunctions \cite{cleland2002noise, cleland2013foundations, paul2006stochastic,clark2010spectral, sader1998frequency}.  The practical aspects of the expansion, however,  become cumbersome with increasing dissipation. Since waves get attenuated along the structure and at the boundaries,  one needs a large, if not infinite, number of terms in the eigenfunction expansion.    Here, we illustrate these complications by examining the position- and frequency-dependent oscillatory dynamics of a NEMS resonator driven by a non-symmetric harmonic force at different dissipation limits.  Instead of  an  expansion including a large number of \textit{undamped} eigenfunctions,  we describe the dynamics efficiently by superposing   waves that are attenuated along the beam. 

\begin{figure*}
    \includegraphics[width=6.75in]{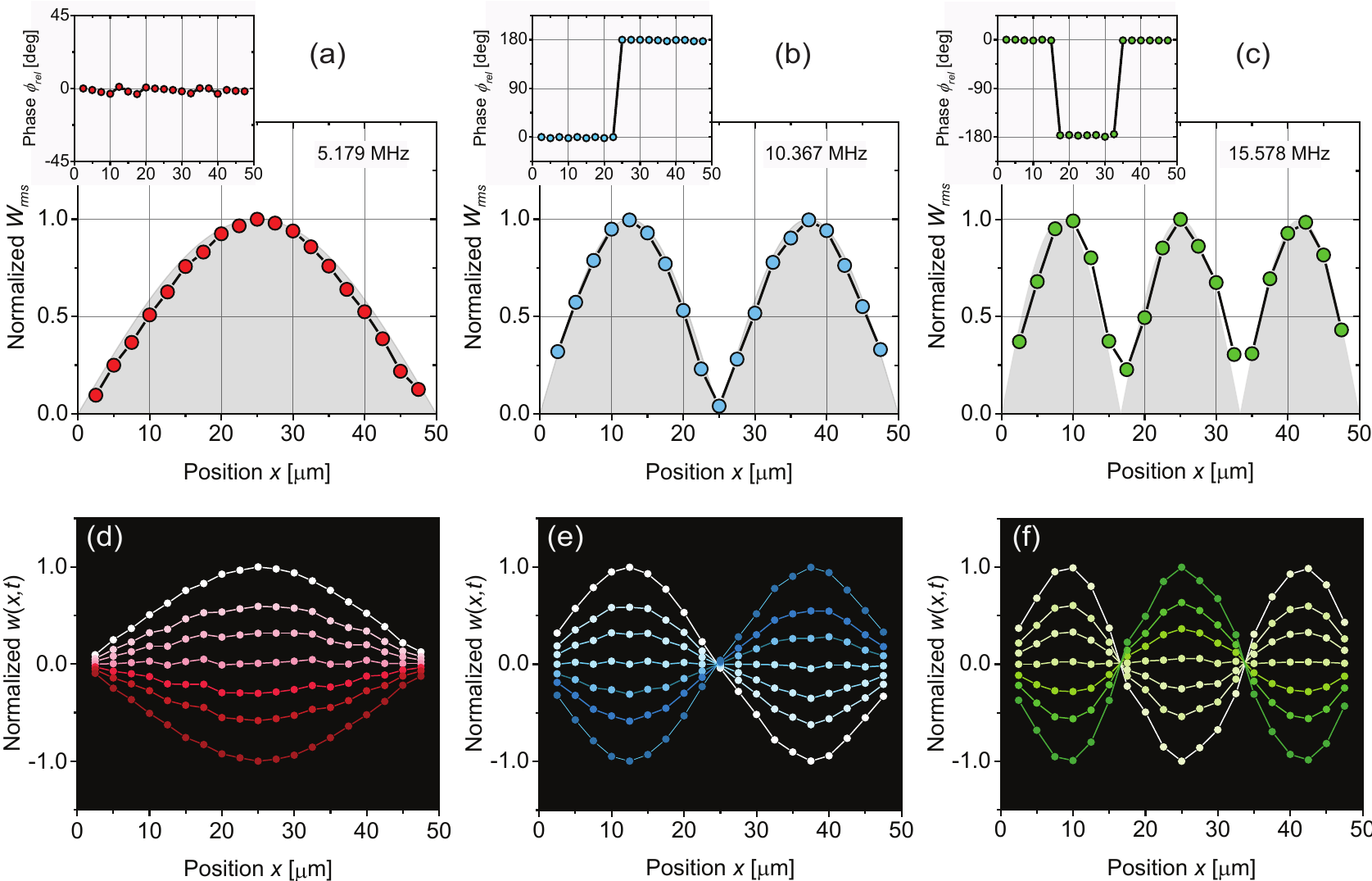}
    \caption{Oscillatory dynamics of a 50-$\rm \mu m $-long  resonator  at the low dissipation limit, i.e., in vacuum. (a-c) Normalized rms amplitudes  $ W_{rms}$ and relative phases $\phi_{rel}$ (insets)  of the beam oscillations, as a function of  $x$ for its first three modes. Data points are from measurements, and the background shadings are $ \left|\sin{(\frac{n \pi}{l}x)}\right|$.  The phase values change by exactly $180^o$ at the nodes.  (d-f) Normalized time-dependent amplitude $w(x,t)$ for the first three modes, as a function of $x$ during an oscillation cycle.  {Relevant parameters of the modes are listed in Table \ref{tab_vacuum}}.}
    \label{Figure2}
\end{figure*}

Our experiments are performed on  silicon nitride doubly-clamped beams with respective linear dimensions along $x$, $y$, and $z$ axes of $l\times b \times h \approx \rm 50~\mu m \times 900~nm \times 100~nm$, all from the same batch. There is a 2-$\rm \mu m$ gap between the beam and the substrate. The beams are under tension as inferred from their resonance frequencies in vacuum \cite{ti2021frequency} and behave as strings. Fig. \ref{Figure1}(a) shows a scanning electron microscope (SEM) image of a beam. The two identical u-shaped gold electrodes are electrothermal actuators for driving the  out-of-plane ($z$-axis) flexural motion of the beam  [Fig. \ref{Figure1}(a) inset]. Each actuator is patterned on one anchor of the resonator with a thickness of 135 nm and a width  of 120 nm. The  actuator spans the undercut  region and  the  beam, with $\xi_1 = 800$ nm and $\xi_2 = 600$ nm; its electrical resistance is $3.54 \pm 0.1~\Omega$ \cite{ti2021frequency}.  We apply  a sinusoidal current  at  frequency $f$ to only one actuator, e.g., the actuator on the left  in Fig. \ref{Figure1}(a).   Joule heating generates temperature oscillations in  the actuator at  $2f$. Owing to the mismatch between the thermal expansion coefficients  of the gold  and  silicon nitride layers, a bending moment develops and drives  out-of-plane flexural oscillations of the beam at $2f$. Fig. \ref{Figure1}(b) shows results from our finite element models of  the electrothermal actuator  in vacuum.  The black curve in Fig. \ref{Figure1}(b)  is the input current waveform at $f = 2.5895~\rm MHz$, with  instantaneous temperature fields [Fig. \ref{Figure1}(b), top]   over the suspended base region of a 50-$\rm \mu m$ resonator.  The red curve in Fig. \ref{Figure1}(b) displays the simulated  amplitude  of the resonator at its center at  $2f = 5.179~\rm MHz$.  For 1 mA  rms input current at $f=2.5895~\rm MHz$, the  temperature oscillates  with rms values of $\Delta T_v=0.1~\rm K$ and $\Delta T_w =0.05~\rm K$ in vacuum and water (not shown), respectively.   The  power dissipated  on the actuator  remains constant  over our frequency range \cite{ti2021frequency,ari2020nanomechanical}. For fixed power, the attenuation of $\Delta T_v$ as a function of frequency is negligible \cite{bargatin2007efficient}; the attenuation of $\Delta T_w$ is expected be even less due to the added thermal conductance of water. Hence, the actuation force is assumed to be independent of frequency.

\begin{figure*}
    \includegraphics[width=6.75in]{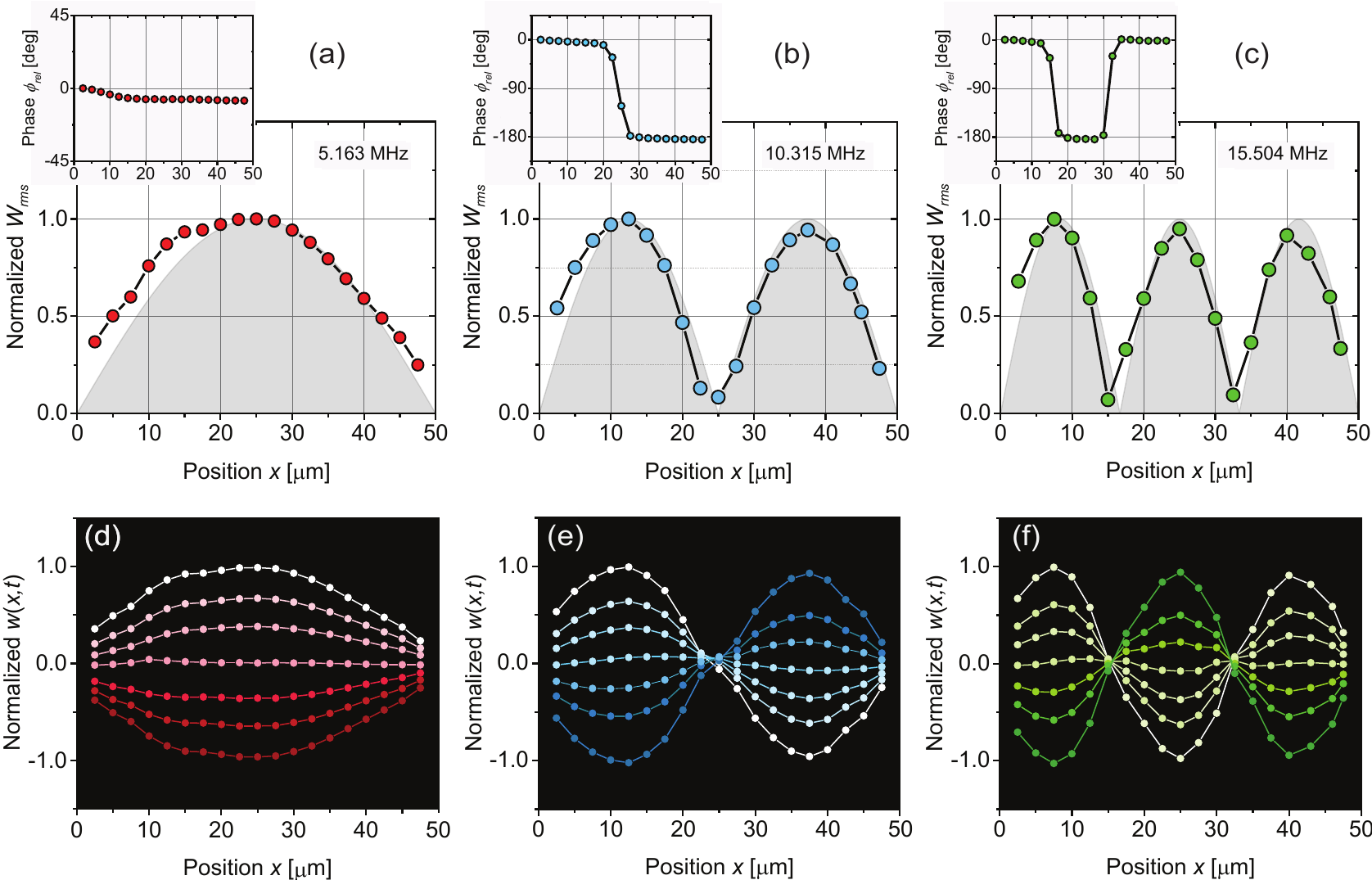}
    \caption{Oscillatory dynamics of the 50-$\rm \mu m $ resonator at intermediate dissipation. (a-c) Normalized rms amplitude  $ W_{rms}$ and relative phase $\phi_{rel}$ (insets)  of the beam, as a function of  $x$ for its first three modes. Data points are from measurements, and the background shadings are $ \left|\sin{(\frac{n \pi}{l}x)}\right|$.   (d-f) Normalized $w(x,t)$ for the first three resonances, as a function of $x$  during the oscillation cycle.   {Relevant parameters of the modes are listed in Table \ref{tab_vacuum}}. }
    \label{Figure3}
\end{figure*}

The harmonically-driven linear dynamics of the NEMS resonator, i.e., the rms amplitude $W_{rms}$ and phase $\phi$ of its oscillations as functions of position,  is measured in a heterodyne optical interferometer \cite{ti2021frequency}. Fig. \ref{Figure1}(c) shows the schematic diagram of the optical  setup \cite{ti2021frequency,wagner1990optical}.  An \textit{XYZ} stage  is used to position the laser spot along the the $x$ axis  in Fig. \ref{Figure1}(a). Fig. \ref{Figure1}(d) shows a representative data set: $W_{rms}$ and  $\phi$  as a function of  $x$ and  drive frequency for the first harmonic mode resonance of the NEMS in vacuum.  The measured $\phi$ can be understood as the phase of the NEMS oscillation with respect to the sinusoidal drive force and includes all the parasitic phases coming from the measurement circuit. We perform the measurements  in vacuum, air, and water,  corresponding respectively to the  low  ($Q\gtrsim 10^4$), intermediate  ($20\lesssim Q \lesssim 70$), and high dissipation ($Q \sim 1$) regimes. In vacuum and air, we measure  the resonator amplitude and phase around the resonance frequencies [Supporting Information Figures S1 and S2]; in water, we sweep the frequency over our entire frequency range.  

\begin{figure*}
    \includegraphics[width=6.75in]{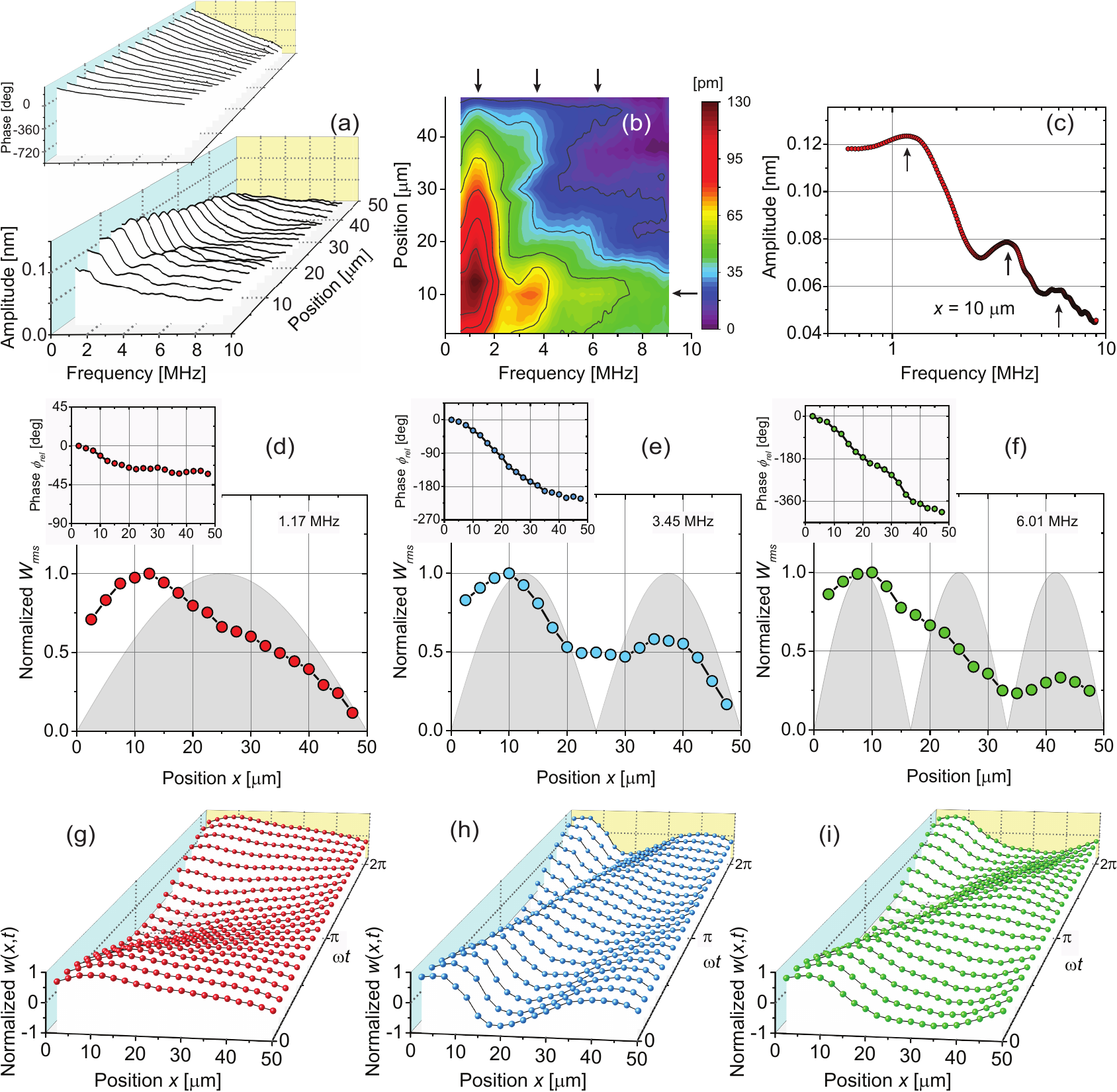}
    \caption{Oscillatory dynamics of the 50-$\rm \mu m $ resonator at  high dissipation. (a)  $W_{rms}$ and phase (inset)  of the  resonator as a function of drive frequency at different $x$ positions on the resonator. (b) Color map of $W_{rms}$. (c)  $W_{rms}$ as a function of frequency  at $x = 10~ \rm \mu m$, with peaks corresponding to the first three resonances.  (d-f) Normalized $W_{rms}$  as a function of $x$  at the peak frequencies in (c) (vertical arrows in (b)): $ 1.17 ~\rm MHz$, $3.45 ~\rm MHz$, and $6.01 ~\rm MHz$. The insets  show  $\phi_{rel}$. (g-i) Normalized $w(x,t)$ constructed from the data in (d-f). A wave propagates from $x\approx 0$ to $x\approx l$ in all cases. The wave decays at different length scales. }    
    \label{Figure4}
\end{figure*}
We first discuss the oscillatory dynamics of the NEMS  at very low dissipation  ($ Q \approx 20 \rm \times 10^3$), as shown in Fig. \ref{Figure2}.  From  measurements  shown in Supporting Information Figure S1, we  obtain the spatial dependence of the rms resonance amplitude $W_{rms}$ and the relative phase $\phi_{rel}$ at each resonance frequency $f_n=\frac{\omega_n}{2\pi}$ [Fig. \ref{Figure2}(a-c)].  $W_{rms}$  are the peak  values  in the amplitude \textit{vs.} frequency curve for each resonance  (Figure S1), and $\phi_{rel}$ is the phase  with respect to the phase of the first data point around $x=0$, i.e., $\phi_{rel}(x)=\phi(x)-\phi ~(x=2.5~\rm \mu m)$, at $f_n$. The time-dependent motion of the beam at frequency $f_n$ can be reconstructed from  $W_{rms}$ and $\phi_{rel}$ as $w(x,t)=\sqrt{2}W_{rms}(x)\sin \left[\omega_n t + \phi_{rel}(x) \right]$ by advancing the dimensionless time $\omega_n t$ over a cycle. Fig. 2(d-f) shows normalized $w(x,t)$ for the first three modes.  Supporting Information includes video files of these modes.

Since $Q$ is very high, we ignore the dissipation and solve the undamped string equation with fixed-fixed boundary conditions to obtain $w(x,t) = \sin{\left(k_{n} x\right)}\sin (\omega_{n} t)$ with $n$ being the mode number \cite{Rao:2007ui}. Here, $k_{n}= n\frac{\pi}{l}$ is the wave number and $f_n =\frac{\omega_n}{2\pi} =n \frac{c}{2l}$ is the corresponding eigen-frequency. The speed of flexural waves in the beam is $c= \sqrt{\frac{\tau}{\mu}} = 510 \pm 10 ~\rm m/s$, based on experimental values\cite{ti2021frequency} of  tension $\tau = 68\pm 4~\rm \mu N$ and mass per unit length $\mu = \rho_{s}bh = 26.6\pm 0.3 \times 10^{-11} ~ \rm kg/m$ with the density being $\rho_{s} = 2960~\pm~30 ~\rm kg/m^{3}~$. The shadings in Figs. \ref{Figure2}(a-c) are based on this solution for $n= 1$, 2 and 3. We note that $w(x,t)$ can also be written as a sum of two undamped propagating waves as $w(x,t) = \Re\{{\frac{1}{2}}{e^{i({k_{n}}x - {\omega _{n}}t)}} - {\frac{1}{2}}{e^{-i({k_{n}}x + {\omega _{n}}t)}}\}$, with $\Re$ denoting the real part of the complex expression. We emphasize that the boundary conditions are not trivial: there are undercuts and the gold nanoresistors around $x = 0$ and $x=l$. The rigidity of the beam also becomes appreciable near the clamps. Regardless, the low dissipation makes these complications negligible. 
\begingroup
\setlength{\tabcolsep}{10pt} 
\renewcommand{\arraystretch}{1} 
\begin{table*}
\caption{ Parameters for the first three modes of the NEMS resonator in vacuum and air: $\omega_{n}/2\pi$ and $Q_{n}$ respectively refer to the mode frequency and quality factor; $k_{n}$ is the mode wave number. Air values are indicated by  $a$; $R$ and $I$ correspond to  real and imaginary components. } \label{tab_vacuum}

\begin{tabular}{cccccccccc}
\hline
& &  & Vacuum &  & &  &   & Air &   \\
\hline
Mode & &  $\frac{\omega_n}{2 \pi}$   & $Q_{n}$ & $k_{n}$ & & $\frac{\omega_{n}^{(a)}}{2 \pi}$ &  $Q_{n}^{(a)}$  & $k_{nR}^{(a)} $ & $k_{nI}^{(a)} $  \\
& & (MHz) &   & ($\rm m^{-1}$) & &  (MHz) & & $(\rm m^{-1})$ &  $(\rm m^{-1}$) \\
\hline
 1 & & 5.179  & 23.6 $\times 10^{3}$ & 6.28 $\times 10^{4}$  & & 5.163 & 32 $\pm~5$ & 6.58 $\times 10^{4}$ & 2.21 $\times 10^{3}$ \\
 2 & & 10.367  & 22.8 $\times 10^{3}$ & 12.57 $\times 10^{4}$ & & 10.315 & 54 $\pm~5$ & 13.18 $\times 10^{4}$& 3.76 $\times 10^{3}$ \\
 3 & &  15.578 & 20.3 $\times 10^{3}$ & 18.85 $\times 10^{4}$ & & 15.504 & 73 $\pm~5$ & 19.76 $\times 10^{4}$ & 4.04 $\times 10^{3}$ \\
 
\end{tabular}
\end{table*}
\endgroup

Now, we turn to the oscillatory dynamics of the same resonator in the intermediate dissipation limit  by repeating the  experiment in air. Since the quality factors, $Q_{n}^{(a)}$, in air are still relatively high  (Table I), the modes are well separated in frequency (Supporting Information Figure S2). At a first glance,  $W_{rms}$ and $\phi_{rel}$   data in Fig. 3(a-c) look similar to those  in Fig. 2. The resonance frequencies also do not deviate much from their vacuum values (Table I).    Upon more careful comparison with Fig. \ref{Figure2}, however,  we notice that the  amplitudes in Fig. 3   become asymmetric with respect to the beam center and decay noticeably  away from the actuator. To  highlight these features, we show in the background of Fig. 3(a-c) the undamped eigenfunctions $ \left|\sin{(\frac{n \pi}{l}x)}\right|$. The step jumps in $\phi_{rel}$  in Fig. \ref{Figure2}(b-c) become smooth in Fig. \ref{Figure3}(b-c), indicating  that the waves are propagating along the $x$. The corresponding  $w(x,t)$ constructed from the data are shown in Fig. 3(d-f). Supporting Information includes video files, where the zero crossings of of $w(x,t)$ move slightly along the $x$.

We  model the  dynamics in air using the  string equation with uniform viscous damping \cite{bokaian1990natural,ti2021frequency,ari2020nanomechanical,liem2020inverse,Rao:2007ui}: 
\begin{equation}
\mu \frac{\partial^2 w}{\partial t^2}  +  \gamma \frac{\partial w}{\partial t} -\tau \frac{\partial^2 w}{\partial x^2}  = f(x,t).
\label{eq: governing}
\end{equation}
Here, $\gamma$ is the damping per unit length and $f(x,t) = \Re\{ F(x) e^{i \omega t}\}$ is the applied force per unit length  with $F(x)$ being the complex force amplitude. Considering only the domain  $\xi_1+\xi_2 < x \le l$ in which $F(x) \approx 0$, we write the general  solution to Eq. \ref{eq: governing} as
\begin{equation}
w(x,t) = \Re\{A e^{ i (k x - \omega t)} + B e^{- i(kx + \omega t)}\}. 
\label{eq:sol}
\end{equation}
The complex wave vector $k$ is found  as
\begin{equation}
k = k_R + ik_I  = \frac{\omega}{c}{\left( {1 + i\frac{\gamma}{{\mu \omega}}  } \right)^{1/2}},
 \label{eq:k}
\end{equation} 
where  $k_{R}=\Re\{k\}$,  $k_I = \Im \{k\} $, and $\Re$ and $\Im$ respectively denote the real and imaginary components. Eq. \ref{eq:sol} can thus  be rewritten as
\begin{equation}
    w(x,t) = \Re\{A e^{-k_I x}e^{i(k_R x  - \omega t)} + B e^{k_I x}e^{-i(k_{R}x + \omega t)}\},
    \label{eq:prop wave}
\end{equation}
with $A$ and $B$ being the complex amplitudes of the right- and left-propagating  waves, respectively. We can fit the data in Fig. \ref{Figure3}(a-c) using Eq. \ref{eq:prop wave}, as shown in Fig. S3. The best fits provide the complex $k$ values listed in Table I. Using $\gamma \approx {\frac{\mu \omega}{Q_{n}^{(a)}} }$, we expand  Eq. \ref{eq:k} to find  ${k_{n}^{(a)}} \approx {\frac{n \pi}{l}}\left( {1 + i{\frac{1}{2Q_{n}^{(a)}}   }} \right)$. The values for $k_{nR}^{(a)}$ are very close to $\frac{n \pi}{l} $;  $k_{nI}^{(a)}$ are roughly a factor of two to three larger than  $ \frac{{n \pi}}{{2 l Q_{n}^{(a)}}} $.    We note that the fits in Fig. S3 are approximations only and can  be improved by  modeling the boundary conditions more realistically.  


\begin{figure}
    \includegraphics[width=3.375in]{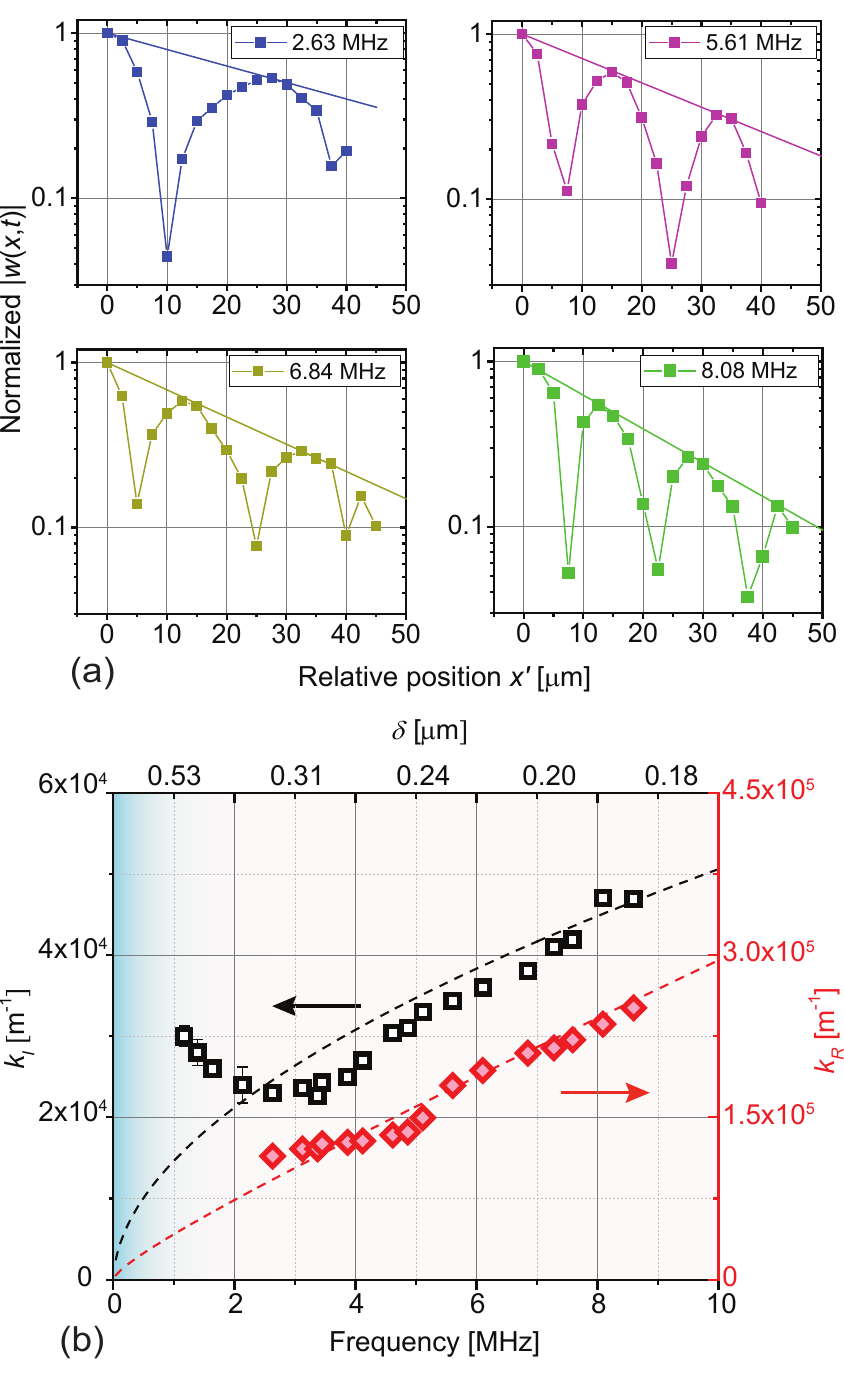} \caption{(a) Semilogarithmic plots showing normalized $\left|w(x,t)\right|$ \textit{vs.} $x'$ at different drive frequencies, where $x'=0$ marks the peak position of the waveform. Fitting the envelope with a decaying exponential (lines) provides $k_{I}$; the distance between successive peaks provides $k_{R}$. (b) $k_{I}$ (left $y$) and $k_{R}$ (right $y$) as a function of frequency. The upper $x$ axis shows $\delta$. The shading indicates the regions $\delta \gtrsim$ gap (dark) and $\delta \lesssim$ gap (light). The dashed curves show theoretical predictions. Error bars are the uncertainties in the linear fits in (a).}
    \label{Figure5}
\end{figure}

Finally, we  show our results on  the oscillatory dynamics of the resonator at the high dissipation limit ($Q \approx 1$) in Fig. \ref{Figure4}. This  experiment is performed with the NEMS immersed in water. Fig. \ref{Figure4}(a) shows  the rms oscillation amplitude and the phase (inset) of the NEMS as a function of frequency and position, obtained by scanning the drive frequency in the $0.6 - 9
\rm~MHz$ range  at each  $x$. The colormap in Fig. \ref{Figure4}(b) is the top view of the  amplitude   from Fig. 4(a), showing how the amplitude decays with frequency and position.  We  observe that the amplitude shows peaks at some frequencies reminiscent of resonances. For instance,  $W_{rms}$  as a function of frequency at $x=10~\rm \mu m$ shown in Fig. \ref{Figure4}(c) has three peaks.  Taking the values of $W_{rms}$ and phase at frequencies marked by the vertical {arrows} in Fig. \ref{Figure4}(b), i.e., at  $ 1.17 ~\rm MHz$, $3.45 ~\rm MHz$, and $6.01 ~\rm MHz$, we obtain the position dependent data for $W_{rms}$ and $\phi_{rel}$ shown in Fig. 4(d-f) at the  peak frequencies in Fig. 4(c).  For more insight into the motion of the beam, we construct $w(x,t)$ as above for  full cycles of oscillation, as shown in Fig. 4(g-i) and Supporting Information videos. Immediately evident is the fact that $w(x,t)$ are traveling waves that are generated at the actuator at $x\approx 0$ and move toward $x=l$.  The waves  decay significantly over the length of the beam. Consequently, we can neglect the  wave propagating to the left in Eq. \ref{eq:sol} and have a simpler mathematical description,
\begin{equation}
\begin{split}
  w(x,t) & \approx \Re\{A e^{ i (k x - \omega t)}\},\\
        &=A'e^{-k_I x} cos(k_R x-\omega t +\varphi),
\end{split}
\label{eq:heavy damped}
\end{equation}
where the phase $\varphi$ is adjusted such that $A'$ is real.  To estimate $k_R$ and  $k_I$ as a function of frequency, we advance the waveform in time (or adjust $\varphi$) until we obtain a peak near the $x=0$ anchor. This results in data sets such as those shown in Fig. 5(a).  In these semilogarithmic graphs, $\left| w(x,t)\right|$ is plotted at four frequencies with the $x'$ axis starting  at the peak position. The distance between successive peaks is half the wavelength and  provides an estimate for $k_R$ of a given data set. The decaying exponential envelope of each data set, shown by the line in each plot, provides an estimate for $k_I$.   In Fig. \ref{Figure5}(b)  the extracted $k_R$ and $k_I$ are plotted  as  functions of  frequency (lower $x$-axis) and the  viscous boundary layer thickness (upper $x$-axis) generated by the oscillations,  $\delta = \sqrt{\frac{2\eta_{f} }{\rho_f \omega}}$, where $\rho_f$ and $\eta_f$ are the density and dynamic viscosity of the  water, respectively.  The $k_R$ data start from  $3~\rm MHz$ since it is impractical to measure the wavelength in the absence of two or more peaks. 

We now show how the frequency-dependent  spatial profile of driven NEMS oscillations can be related to the physical properties of the fluid by using  Stokes' theory of the oscillating cylinder in a viscous fluid \cite{clark2010spectral, sader1998frequency, liem2021nanoflows}.  We Fourier transform  the  undamped string equation ($\gamma=0$ and $f(x,t)=0$  in Eq. \ref{eq: governing}) in both space and time  with the fluid providing the only force (per unit length), ${\tilde F_f}(k, \omega)$, on the beam:
\begin{equation}
    (-\omega^2 \mu + \tau k^2) {\tilde W}(k, \omega)={\tilde F_f}(k, \omega).
    \label{eq:water gov}
\end{equation}
The fluid force is \cite{clark2010spectral} 
\begin{equation}
\begin{split}
    {\tilde F_f}(k, \omega) & \approx {\frac{\pi}{4}} \rho_f \omega^2 b^2 \Gamma_b(\omega){\tilde W}(k, \omega), \\
    & \approx \mu \omega^2 T_0   \Gamma_b(\omega){\tilde W}(k, \omega),
    \end{split}
\end{equation} 
where  $\Gamma_b ({\omega}) = {\Gamma_{b}^{'}} ({\omega}) +  i {\Gamma_{b}^{''}} ({\omega})$ is the  hydrodynamic function of a blade (found from the cylinder solution \cite{sader1998frequency}); $\mu=\rho_s bh$ is the mass per unit length of the string; and $\rho_f$ and  $\eta_f$ are respectively subsumed into  $T_{0} =  \frac{\pi}{4} \frac{\rho_f b}{\rho_s h}$ and $\Gamma_b(\omega)$. Substituting ${\tilde F_f}(k, \omega)$ into Eq. \ref{eq:water gov}, we  obtain
\begin{equation}
k_{R} + i k_{I} = \frac{\omega}{c}\sqrt{{ \left(1 +   T_{0}{\Gamma_{b}^{'}} ({\omega})\right) + i   T_{0} {\Gamma_{b}^{''}} ({\omega})} }. 
\label{eq:k hydro}
\end{equation}
indicating that one can determine $\rho_f$ and $\eta_f$ from  measured  $k_R$ and $k_I$. This approach could be complementary to that based on fitting the frequency response of the NEMS at a single point. In fact, this theory can be extended to obtain properties of viscoelastic fluids as well \cite{hopkins2016vibrating,malara2017rheology}.

The dashed line in Fig. 5(b) show  $k_R$ and $k_I$  predicted  from Eq. \ref{eq:k hydro} using experimental  values of $c$ and $T_0$ and  calculated $\Gamma_b(\omega)$.  These predictions  match well with our experimental data for frequencies$ \gtrsim 2.5~ \rm MHz$. The measured $k_I$ deviates from theory  at low frequency  because of the added squeeze damping \cite{liem2021nanoflows}.  
 
As more emphasis is put on precision measurements in fluids \cite{erdogan2022atmospheric, roy2018improving}, the spatial decay of the amplitude of driven NEMS resonators could have significant implications. For NEMS-based mass sensing and mass spectrometry, deconvoluting the mass and  position of  the adsorbed analyte molecule on the NEMS  from frequency shifts requires a detailed knowledge of the oscillatory amplitude of the resonator in multiple modes \cite{hanay2012single,hanay2015inertial}.  Of particular importance is the behavior of the nodes in the intermediate dissipation regime: since there is a  travelling wave along the structure with a small amplitude, there are no true nodes. Similarly, in dynamic AFM   in air and  liquids \cite{raman2011mapping},  where the microcantilever is driven should affect the tip amplitude  and  tip-sample interactions. Another  relevant area is fundamental studies in fluid dynamics using NEMS and microcantilever resonators \cite{ari2020nanomechanical,clark2010spectral}. The accuracy of an eigenfunction expansion containing  a few eigenmodes  should be assessed carefully in liquids \cite{clark2010spectral,ari2020nanomechanical}. In summary, our results here   will be of  relevance to  research and technology involving NEMS, AFM, and even macroscopic mechanical resonators.



\section*{Supplementary Material}
See the supplementary material for three figures and nine videos. 

\begin{acknowledgments}
We acknowledge support from the US NSF (CBET 1604075, CMMI 1934271, CMMI 2001403, DMR 1709282, and CMMI 1661700). We thank Atakan B. Ari for help with sample fabrication.
\end{acknowledgments}

\section*{AUTHOR DECLARATIONS}
\begin{flushleft}
\textbf{Conflict of Interest}
\end{flushleft}

The authors have no conflicts to disclose.
\begin{flushleft}
\textbf{Author Contributions}
\end{flushleft}

K.L.E., C.T., and M.G. devised the project. C.T. performed the experiments and analyzed the data. C.Y., I.I.K. and M.S.H. fabricated the samples. G.M., C.T., and A.L. developed the wave model and analyzed the data.  M.M. and S.K. performed the  FEM analysis. H.G. characterized the samples. O.S. contributed to the experimental set up.  C.T. and K.L.E. wrote the manuscript with comments from all authors.
\section*{DATA AVAILABILITY}
The data that support the findings of this study are available
from the corresponding authors upon reasonable request.

\bibliography{Reference.bib}

\setcounter{figure}{0}
\renewcommand{\figurename}{Fig.}
\renewcommand{\thefigure}{S\arabic{figure}}
\newpage

\end{document}


\title{Supplementary Figures for \\ ``Dynamics of  NEMS Resonators across Dissipation Limits"}

\newcommand{\BU}{Department of Mechanical Engineering, Division of Materials Science and Engineering, and the Photonics Center, Boston University, Boston, Massachusetts 02215, USA}

\author{C. Ti}
\affiliation{\BU}

\author{J. G. McDaniel}
\affiliation{\BU}

\author{A. Liem}
\affiliation{\BU}

\author{H. Gress}
\affiliation{\BU}

\author{M. Ma}
\affiliation{\BU}

\author{S. Kyoung}
\affiliation{\BU}

\author{O. Svitelskiy}
\affiliation{Department of Physics, Gordon College, Wenham, Massachusetts 01984, United States}

\author{C. Yanik}
\affiliation{SUNUM, Nanotechnology Research and Application Center, Sabanci University, Istanbul, 34956, Turkey}

\author{I. I. Kaya}
\affiliation{Faculty of Engineering and Natural Sciences, Sabanci University, Istanbul, 34956, Turkey}
\affiliation{SUNUM, Nanotechnology Research and Application Center, Sabanci University, Istanbul, 34956, Turkey}

\author{M. S. Hanay}
\affiliation{Department of Mechanical Engineering, Bilkent University, Ankara, 06800, Turkey}
\affiliation{National Nanotechnology Research Center (UNAM), Bilkent University, 06800, Ankara, Turkey}

\author{M. Gonz\'alez}
\affiliation{Aramco Americas, Aramco Research Center--Houston, Houston, Texas, 77084, United States}

\author{K. L. Ekinci}
\email[Electronic mail: ]{ekinci@bu.edu}
\affiliation{\BU}

\date{\today}

\maketitle


\newpage

\widetext

\setcounter{equation}{0}
\setcounter{figure}{0}
\setcounter{table}{0}
\setcounter{page}{1}
\makeatletter
\renewcommand{\theequation}{S\arabic{equation}}
\renewcommand{\thefigure}{S\arabic{figure}}
\renewcommand{\bibnumfmt}[1]{[S#1]}
\renewcommand{\citenumfont}[1]{S#1}

\maketitle

\begin{figure*}
   \includegraphics[width=6.75in]{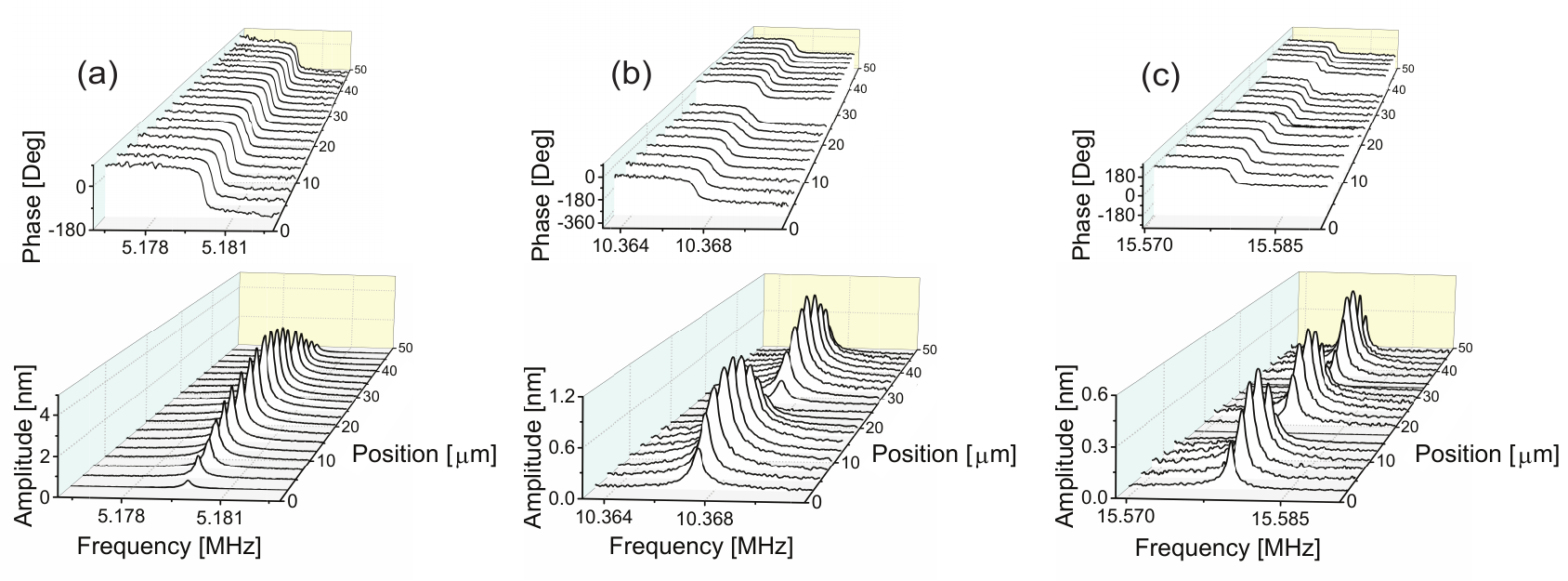}
    \caption{(a-c) Vacuum data showing  the rms amplitude and phase  of the  50-$\rm \mu m $ resonator as a function of frequency around its first three resonances at different positions. }
    \label{FigureS1}
\end{figure*}
\begin{figure*}
    \includegraphics[width=6.75in]{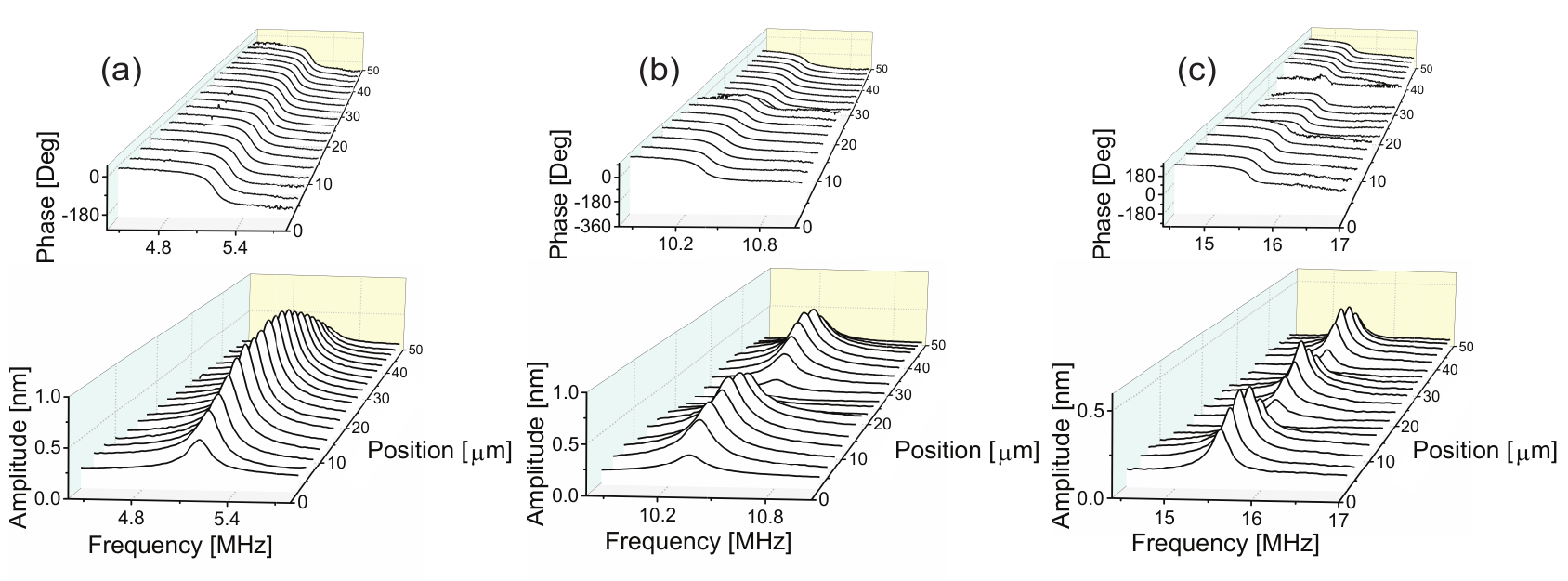}
    \caption{(a-c) Air data showing  the rms amplitude and phase  of the  50-$\rm \mu m $ resonator as a function of frequency around its first three resonances at different   positions. }
    \label{FigureS2}
\end{figure*}

\begin{figure*}
    \includegraphics[width=6.75in]{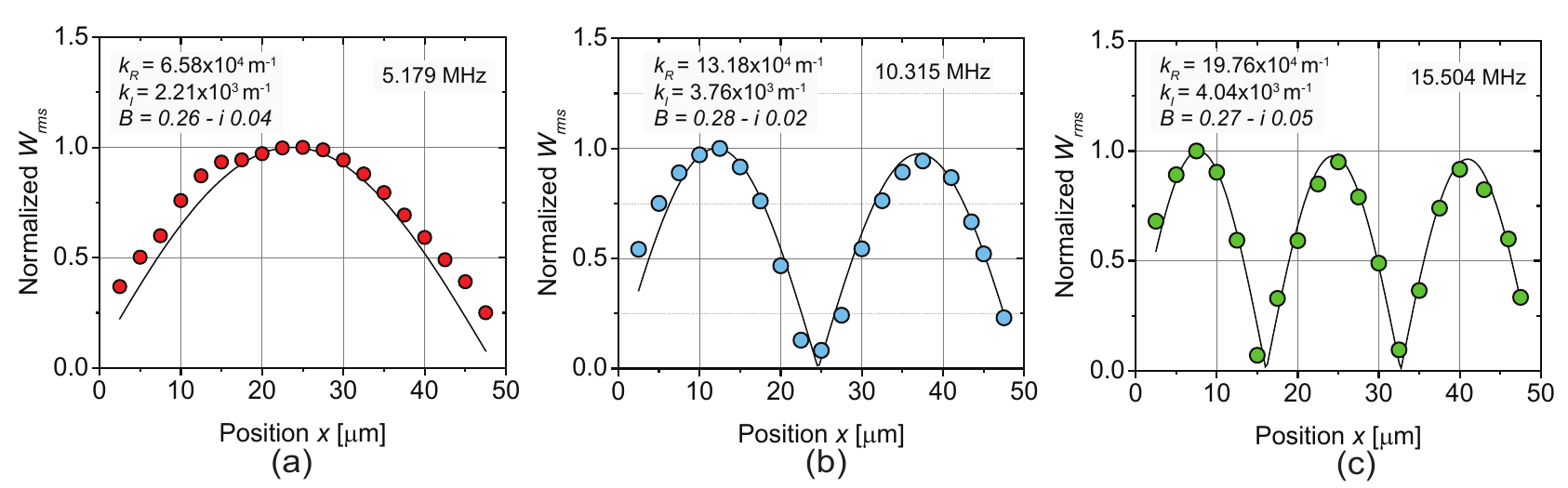}
    \caption{(a-c) Air data and wave fits (continuous curves). Here, the wave is of the form $w(x,t) = \Re\{A e^{-k_I x}e^{i(k_R x  - \omega t)} + B e^{k_I x}e^{-i(k_{R}x + \omega t)}\}$. To make the fits, the $W_{rms}$ and $\phi_{rel}$ data was used to construct the wave at an instant in time ($\omega t= {\pi \over 2}$). Since the dissipation is small, the real component of the wave was determined using the resonance frequency $\omega_n$ as $k_{R} \approx \frac{\omega_n}{c}$.  After fixing $A$ to  $A=0.5$, a search was performed to find the set of  parameters $k_I$ and $B$ that provided the best fits: the error between the normalized experimental wave data  and the wave $w(x,t) = \Re\{A e^{-k_I x}e^{i(k_R x  - \omega t)} + B e^{k_I x}e^{-i(k_{R}x + \omega t)}\}$, for each mode, was minimized.  The found parameters are noted on the plots.}
    \label{FigureS2}
\end{figure*}
